# A Simple Mathematical Model of Politics (II)

## Joey Huang

In previous paper [1], the political status of a society is represented by an n-by-n matrix and it indicates how much people listen to each other. This kind of matrix has been used in many fields – in general it's called a stochastic matrix. The stochastic matrix was first developed by Andrey Markov in 1906 so it's also called a Markov matrix [2]. In social network theory, Morris DeGroot [3] defined a trust matrix for people who "must" act together as a team or committee and described how people could reach a possible consensus based on that – it's called DeGroot Learning. In [1], people do reach a consensus – when selfishness of human beings is ignored. In this case the politics matrix $A$ is like the trust matrix of Morris DeGroot. When people are assumed to be selfish, it leads to more complicated behavior of human beings and the motivation to form some families.

Unlike the trust matrix, every component of $A$ is requested to be positive. However, a directed graph [4] can also be defined by the dominated politics matrix $\hat{A}$. $\hat{A}$ defines a topology structure and a family is an "open set".

It's shown in [1] that a society $\{1,2,...n\}$ is the disjoint sum of some upper-class families ($U = \bigcup_{j=1}^{q} U_j$) and "low-class people" ($L = U'$). It doesn't mean a society is always "socially unfair" – $L$ can be an empty set.

Furthermore, it doesn't mean a society is always divided, either – $q$ can be equal to one (We Are the World). However, when all the people are selfish, this kind of society is not stable.

In this paper, some main eigenvalues and eigenvectors of the politics matrix $A$ are investigated. The number of upper-class families in a society is the number of eigenvalues which are very close to 1. An algorithm to identify all the upper-class families from the right and left eigenvectors of those eigenvalues is developed.

## I. Family Normal Form and Its Main Eigenvalues/Eigenvectors

In this section, $q$ main eigenvalues and eigenvectors are observed in a society with $q$ upper-class families. There are from a multiple root of $det(\lambda I - \hat{A})$: 1.

When people in the same family "group together", the dominated politics matrix is written in this "family normal form":

$$\hat{A} = \begin{bmatrix} \hat{A}_{UU} & 0 \\ \hat{A}_{LU} & \hat{A}_{LL} \end{bmatrix}$$

$$\hat{A}_{UU} = \begin{bmatrix} \hat{A}_{11} & \cdots & 0 \\ \vdots & \ddots & \vdots \\ 0 & \cdots & \hat{A}_{qq} \end{bmatrix}$$

$$A = \hat{A} + \varepsilon B, \hat{A}\eta = \eta, B\eta = 0$$

$\hat{A}_{jj}$ means $\hat{A}_{U_jU_j}$.

The characteristic polynomial of $\hat{A}$ is

$$det(\lambda I - \hat{A}) = det(\lambda I_L - \hat{A}_{LL}) \cdot \prod_{j=1}^{q} det(\lambda I_j - \hat{A}_{jj})$$

By Lemma 7 in [1], $rank(I_j - \hat{A}_{jj}) = m_j - 1$, $m_j$ is the number of people in $U_j$. So 1 is a simple root of $det(\lambda I_j - \hat{A}_{jj})$ and the "*internal power*" $\varpi_j^*$ of upper-class family $U_j$ can be found:

$$\varpi_j^* \hat{A}_{jj} = \varpi_j^*, \varpi_j^* \eta_j = 1, \varpi_j^* > 0$$

$\varpi_j^* > 0$ means every member of this upper-class family gets a meaningful slice of internal power – big or small.

$I_L - \hat{A}_{LL}$ is invertible by Lemma 6 in [1] so 1 is not a root of $det(\lambda I_L - \hat{A}_{LL})$. In total 1 is a multiple root of $det(\lambda I - \hat{A})$ and $q$ is the multiplicity:

*The multiplicity of eigenvalue* 1 *indicates the number of upper-class families in a society.*

The q-by-n matrix $W^*$ is defined by

$$W_U^* = \begin{bmatrix} \varpi_1^* & \cdots & 0 \\ \vdots & \ddots & \vdots \\ 0 & \cdots & \varpi_q^* \end{bmatrix}, W^* \equiv [W_U^* \quad 0_{q \times l}]$$

$l$ is the number of people in $L$.

$W^* \hat{A} = W^*$ and the $q$ rows of $W^*$ form a basis of the vector space $\{u^*: u^* \hat{A} = u^*\}$.

The $q$ columns of this n-by-q matrix $J$ form a basis of the vector space $\{v: \hat{A}v = v\}$ – it's more complicated than $W^*$:

$$J_U = \begin{bmatrix} \eta_1 & \cdots & 0 \\ \vdots & \ddots & \vdots \\ 0 & \cdots & \eta_q \end{bmatrix}, J = \begin{bmatrix} J_U \\ J_L \end{bmatrix} \equiv \begin{bmatrix} J_U \\ \widehat{D}_{LU}J_U \end{bmatrix} = \begin{bmatrix} J_U \\ (I - \hat{A}_{LL})^{-1}\hat{A}_{LU}J_U \end{bmatrix}$$

$$\hat{A}J = J, J\boxed{\eta} = \eta$$

$\eta_j$ is the $m_j$-by-1 all-ones vector and $\boxed{\eta}$ is the q-by-1 all-ones vector.

The l-by-q matrix $J_L = \widehat{D}_{LU} J_U$ indicates how much those lower-class people are "with" each upper-class family. If there are no low-class people in this society, $J$ is also very simple like $W^*$. The whole n-by-q matrix $J$ indicates how much every individual is with each upper-class family – an upper-class person is with his upper-class family, 100%.

$W^*$ and $J$ are "perfectly orthogonal":

$$W^* J = I$$

Now let's look at the "real" politics matrix $A$ instead of the dominated politics matrix $\hat{A}$. 1 is just a simple eigenvalue of $A$ and there should be $q$ eigenvalues close to 1 – including 1 itself:

$$\lambda = I + \varepsilon \dot{\lambda} + O(\varepsilon^2)$$

$\dot{\lambda}$ is an q-by-q *diagonal* matrix.

Let $V = \hat{V} + \varepsilon \dot{V} + O(\varepsilon^2)$ be an n-by-q matrix representing $q$ eigenvectors of $A$:

$$AV = V\lambda$$

$$(\hat{A} + \varepsilon B)(\hat{V} + \varepsilon \dot{V}) = (\hat{V} + \varepsilon \dot{V})(I + \varepsilon \dot{\lambda}) + O(\varepsilon^2)$$

$$\hat{A}\hat{V} = \hat{V}$$

$$B\hat{V} + \hat{A}\dot{V} = \hat{V}\dot{\lambda} + \dot{V}$$

Let's normalize $V$ by $diag(V^*V) = I$ and the first column $v_1 = \eta/\sqrt{n}$ is the trivial eigenvector of $A$ for $\lambda_1 = 1 + \varepsilon \dot{\lambda}_1 = 1$.

Every column of $\hat{V}$ is a linear combination of the basis $J$ so $\hat{V} = J\beta$, $\beta$ is an q-by-q matrix.

$$BJ\beta + \hat{A}\dot{V} = J\beta\dot{\lambda} + \dot{V}$$

(1)

$\eta/\sqrt{n} = v_1 = \hat{v}_1 = J\beta_1$. $\beta_1 = \boxed{\eta}/\sqrt{n}$.

So every eigenvector is almost like a step function for upper-class people. A "step" represents a family – don't forget $\varepsilon$ is still there so it won't be a "perfect step".

$diag(\hat{V}^* \hat{V}) = I$ so

$$I = diag(\beta^* J^* J \beta) = diag(\beta^*(J_U^* J_U + J_L^* J_L)\beta)$$

$$J_U^* J_U = \begin{bmatrix} m_1 & \cdots & 0 \\ \vdots & \ddots & \vdots \\ 0 & \cdots & m_q \end{bmatrix}$$

On the other hand, let $U^* = \hat{U}^* + \varepsilon \dot{U}^* + O(\varepsilon^2)$ be an q-by-n matrix representing $q$ left eigenvectors of $A$:

$$U^* A = \lambda U^*$$

$$(\hat{U}^* + \varepsilon \dot{U}^*)(\hat{A} + \varepsilon B) = (I + \varepsilon \dot{\lambda})(\hat{U}^* + \varepsilon \dot{U}^*) + O(\varepsilon^2)$$

$$\hat{U}^* \hat{A} = \hat{U}^*$$

$$\hat{U}^* B + \dot{U}^* \hat{A} = \dot{\lambda} \hat{U}^* + \dot{U}^*$$

Let's normalize $U^*$ by $U^* V = I$. The power $\omega^*$ of $A$ appears in the first row $u_1^* = \sqrt{n} \cdot \omega^*$.

Similarly, $\hat{U}^* = \alpha^* W^*$, $\alpha^*$ is an q-by-q matrix.

$$\alpha^* W^* B + \dot{U}^* \hat{A} = \dot{\lambda} \alpha^* W^* + \dot{U}^*$$

(2)

The dominated power is determined by $\alpha_1^*$:

$$\hat{\omega}^* = \frac{\hat{u}_1^*}{\sqrt{n}} = \frac{\alpha_1^* W^*}{\sqrt{n}} = \frac{1}{\sqrt{n}} \left[ \alpha_{11}^* \varpi_1^* \; \alpha_{12}^* \varpi_2^* \; \ldots \; \alpha_{1q}^* \varpi_q^* \; 0_{1 \times l} \right]$$

Low-class people don't get any dominated power – it has been proved in [1].

$\alpha^*$ is the inverse of $\beta$:

$$I = \hat{U}^* \hat{V} = \alpha^* W^* J \beta = \alpha^* \beta$$

Multiply both sides of (1) by " $W^* \cdot$ " and both sides of (2) by " $\cdot J$ ",

$$(W^* B J) \beta = \beta \dot{\lambda}, \; \alpha^* (W^* B J) = \dot{\lambda} \alpha^*$$

(3)

So $\dot{\lambda}$, $\beta$ and $\alpha^*$ serve as all the eigenvalues and eigenvectors of this q-by-q matrix $W^* B J$. Let's create a cyber society with $q$ members based on this. The politics matrix of this cyber society is

$$\boxed{A} = I + \varepsilon \boxed{B} = W^* A J = I + \varepsilon (W^* B J)$$

A member of this cyber society represents an upper-class family of the real society. This cyber society is like an almost-no-politics society – because upper-class people don't "play meaningful politics" with people out of their families. It's easy to see that $I + \varepsilon \dot{\lambda}$ represents all the eigenvalues of $\boxed{A}$, $\alpha^*$ and $\beta$ represent all the left and right eigenvectors of $\boxed{A}$. *The political structure of this cyber society with q members plus $W^*$ and J determine the q main eigenvalues and eigenvectors of A.*

Lower-class people are "invisible" in this cyber society, unfortunately. They play a part in the structure of this cyber society by the second term of this formula:

$$\boxed{B} = W^*BJ = [W_U^* \quad 0]\begin{bmatrix} B_{UU} & B_{UL} \\ * & * \end{bmatrix}\begin{bmatrix} J_U \\ \widehat{D}_{LU}J_U \end{bmatrix} = W_U^*B_{UU}J_U + W_U^*B_{UL}\widehat{D}_{LU}J_U$$

$$\boxed{B}_{ij} = \varpi_i^* B_{ij} \eta_j + \varpi_i^* B_{iL} \widehat{D}_{Lj} \eta_j$$

When $i \neq j$, how much person $i$ listens to person $j$ slightly in this cyberspace is the weighted average of how much people in upper-class family $U_i$ listen to people in upper-class family $U_j$ slightly ($\varepsilon \varpi_i^* B_{ij} \eta_j$) plus the weighted average of how much they listen to low-class people times how much those low-class people are with $U_j$ ($\varepsilon \varpi_i^* B_{iL} \widehat{D}_{Lj} \eta_j$). Low-class people play the role as "extra effect" in this cyber society.

$\alpha_1^* = \sqrt{n} \cdot \boxed{\varpi}^*$ so that $\alpha_1^* \beta_1 = \sqrt{n} \cdot \boxed{\varpi}^* \boxed{\eta}/\sqrt{n} = 1$. $\boxed{\varpi}^*$ is the power of this cyber society.

$$\widehat{\omega}^* = \boxed{\varpi}^* W^* = \begin{bmatrix} \boxed{\varpi}_1^* \varpi_1^* & \boxed{\varpi}_2^* \varpi_2^* & \cdots & \boxed{\varpi}_q^* \varpi_q^* & 0_{1 \times l} \end{bmatrix}$$

As expected, $\widehat{\omega}^* \eta = \boxed{\varpi}^* W^* \eta = \boxed{\varpi}^* \boxed{\eta} = 1$.

So, the political structure of this cyber society determines the main political structure of the original society. Don't forget there are still other small $n - q$ eigenvalues of $A$ and $\varepsilon$ is still there.

## II. The Next-Order Terms of Main Eigenvalues/Eigenvectors

It's more complicated to get the next-order terms. They can help to pick up a right diagonal matrix for the algorithm in next section.

$\dot{\lambda}$, $\alpha^*$ and $\beta$ can be obtained from the q-by-q matrix $W^*BJ$ in Equation (3). Now Equation (1) is for the unknown $\dot{V}$ and Equation (2) is for the unknown $\dot{U}^*$:

$$(I - \widehat{A})\dot{V} = BJ\beta - J\beta\dot{\lambda}$$

$$\dot{U}^*(I - \widehat{A}) = \alpha^* W^* B - \dot{\lambda}\alpha^* W^*$$

Let's introduce this n-by-n matrix $S$ defined by the politics matrix $A$:

$$S \equiv \lim_{\varepsilon \to 0} \left((1+\varepsilon)I - A\right)^{-1} - \frac{1}{\varepsilon}\eta\omega^*$$

(4)

Because $I - A$ is singular, $\varepsilon$ is there to make $(1+\varepsilon)I - A$ invertible. There's an explicit formula for $S$ (see Lemma 9 in [1]):

$$S = (I - \eta\omega^*)Z(I - \eta\omega^*)$$

$$Z = \begin{bmatrix} (I - A_{GG})^{-1} & 0 \\ 0 & 0 \end{bmatrix}, G = \{1,2,\ldots n-1\}$$

$$S\eta = 0, \omega^* S = 0$$

There's another formula of $S$ – it can be the definition of $S$ as well:

$$S = \lim_{\varepsilon \to 0} \frac{1}{1+\varepsilon}\left(I - \frac{A}{1+\varepsilon}\right)^{-1} - \frac{1}{1+\varepsilon}\sum_{i=0}^{\infty} \frac{\eta\omega^*}{(1+\varepsilon)^i}$$

$$S = \lim_{\varepsilon \to 0} \frac{1}{1+\varepsilon}\sum_{i=0}^{\infty} \frac{A^i}{(1+\varepsilon)^i} - \frac{1}{1+\varepsilon}\sum_{i=0}^{\infty} \frac{A^\infty}{(1+\varepsilon)^i}$$

$$S = \sum_{i=0}^{\infty} \left(A^i - A^\infty\right)$$

It' clear that $S$ commutes with $A$, $SA = AS = S - (I - A^\infty)$. So,

$$S(I - A) = (I - A)S = I - \eta\omega^*$$

$S$ serves as the "inverse" of $I - A$ when it's not invertible. It appears in the solution of this equation:

$$(I - A)x = b$$

First, $\omega^* b = \omega^*(I - A)x = 0$ – otherwise there's no solution. Then $x = Sb$ is a special solution because $(I - A)x = (I - A)Sb = (I - \eta\omega^*)b = b$. The general solution is $x = Sb + c\eta$ for any constant $c$.

Similarly, $x^* = b^*S + c\omega^*$ is the general solution of $x^*(I - A) = b^*$ for any $c$ when $b^*\eta = 0$.

However, $I - \hat{A}$ is "more singular" than $I - A$ so the definition of $\hat{S}$ is more complicated:

$$\hat{S} = \begin{bmatrix} \hat{S}_U & 0 \\ \hat{D}_{LU}\hat{S}_U & (I - \hat{A}_{LL})^{-1} \end{bmatrix}$$

$$\hat{S}_U = \begin{bmatrix} \hat{S}_1 & \cdots & 0 \\ \vdots & \ddots & \vdots \\ 0 & \cdots & \hat{S}_q \end{bmatrix}$$

For any $j$, $\hat{S}_j$ is defined by $\hat{A}_{jj}$ based on Equation (4) when $rank(I_j - \hat{A}_{jj}) = m_j - 1$.

$$\hat{S}_j \eta_j = 0, \varpi_j^* \hat{S}_j = 0$$

$$\hat{S}_j(I - \hat{A}_{jj}) = (I - \hat{A}_{jj})\hat{S}_j = I - \eta_j \varpi_j^*$$

$$\hat{S}_U J_U = 0, W_U^* \hat{S}_U = 0$$

$$\hat{S}_U(I - \hat{A}_{UU}) = (I - \hat{A}_{UU})\hat{S}_U = I - J_U W_U^*$$

$$\hat{S}J = \begin{bmatrix} 0 \\ (I - \hat{A}_{LL})^{-1} J_L \end{bmatrix} = \begin{bmatrix} 0 \\ (I - \hat{A}_{LL})^{-2} \hat{A}_{LU} J_U \end{bmatrix}, W^*\hat{S} = 0$$

$\hat{S}J = 0$ when there are no low-class people in this society.

Special solutions of $\dot{V}$ and $\dot{U}^*$ are

$$\dot{V}_0 = \hat{S}(BJ\beta - J\beta\lambda), W^*\dot{V}_0 = 0$$

$$\dot{U}_0 = (\alpha^*W^*B - \lambda\alpha^*W^*)\hat{S} = \alpha^*W^*B\hat{S}$$

$J$ and $W^*$ play the roles as bases again for general solutions of $\dot{V}$ and $\dot{U}^*$. For convenience they are replaced by $\hat{V} = J\beta$ and $\hat{U}^* = \alpha^*W^*$:

$$\dot{V} = \dot{V}_0 + \hat{V}\dot{\beta} = \dot{V}_0 + J\beta\dot{\beta}$$

$$\dot{U}^* = \dot{U}_0 + \dot{\alpha}^*\hat{U}^* = \dot{U}_0 + \dot{\alpha}^*\alpha^*W^*$$

It's more complicated to get both $\dot{\alpha}^*$ and $\dot{\beta}$. Let's look at the Taylor series with one more term:

$$(\hat{A} + \varepsilon B)(\hat{V} + \varepsilon \dot{V} + \varepsilon^2 \ddot{V}) = (\hat{V} + \varepsilon \dot{V} + \varepsilon^2 \ddot{V})(I + \varepsilon \dot{\lambda} + \varepsilon^2 \ddot{\lambda}) + O(\varepsilon^3)$$

$$B\dot{V} + \hat{A}\ddot{V} = \hat{V}\ddot{\lambda} + \dot{V}\dot{\lambda} + \ddot{V}$$

$\ddot{V}$ disappears when both sides are multiplied by " $W^* \cdot$ ":

$$W^*B\dot{V}_0 + (W^*BJ)\beta\dot{\beta} = \beta\ddot{\lambda} + \beta\dot{\beta}\dot{\lambda}$$

The q-by-q matrix $\boxed{B} = W^*BJ$ appears again. Multiply both sides by "$\alpha^* \cdot$", $\alpha^*(W^*BJ) = \dot{\lambda}\alpha^*$:

$$\alpha^*W^*B\dot{V}_0 + \dot{\lambda}\dot{\beta} = \ddot{\lambda} + \dot{\beta}\dot{\lambda}$$

Now the q-by-q matrix $\dot{\beta}$ is decomposed into the diagonal part and the rest:

$$\dot{\beta}_d = diag(\dot{\beta}), \dot{\beta} = \dot{\beta}_d + \dot{\beta}_0$$

Because $\dot{\lambda}$ is diagonal, $\dot{\lambda}\dot{\beta}_d = \dot{\beta}_d\dot{\lambda}$,

$$\alpha^*W^*B\dot{V}_0 - \ddot{\lambda} = \dot{\beta}_0\dot{\lambda} - \dot{\lambda}\dot{\beta}_0$$

Even though $\ddot{\lambda}$ is unknown, it's diagonal. So every component of $\dot{\beta}_0$ can be obtained by

$$(\alpha^*W^*B\dot{V}_0)_{ij} = (\dot{\lambda}_j - \dot{\lambda}_i)(\dot{\beta}_0)_{ij} \text{ when } i \neq j$$

$\ddot{\lambda}$ can be obtained by $\ddot{\lambda} = diag(\alpha^*W^*B\dot{V}_0)$.

Now let's look at the equation to normalize $V$:

$$I = diag(V^*V) = diag(\hat{V}^*\hat{V}) + 2\varepsilon \cdot diag(\hat{V}^*\dot{V}) + O(\varepsilon^2)$$

$$0 = diag(\hat{V}^*\dot{V}) = diag(\hat{V}^*\dot{V}_0) + diag(\hat{V}^*\hat{V}\dot{\beta}_0) + diag(\hat{V}^*\hat{V}\dot{\beta}_d)$$

Because $\dot{\beta}_d$ is diagonal, $diag(\hat{V}^*\hat{V}\dot{\beta}_d) = diag(\hat{V}^*\hat{V}) \cdot \dot{\beta}_d = \dot{\beta}_d$

$$\dot{\beta}_d = -diag\left(\hat{V}^*(\dot{V}_0 + \hat{V}\dot{\beta}_0)\right) = -diag\left(\beta^*J^*(\dot{V}_0 + J\beta\dot{\beta}_0)\right)$$

$\dot{\alpha}^*$ can be obtained easily from the equation to normalize $U^*$:

$$I = U^*V = \hat{U}^*\hat{V} + \varepsilon\dot{U}^*\hat{V} + \varepsilon\hat{U}^*\dot{V} + O(\varepsilon^2)$$

$$0 = \dot{U}^*\hat{V} + \hat{U}^*\dot{V} = \dot{U}_0\hat{V} + \dot{\alpha}^* + \dot{\beta}$$

$$\dot{\alpha}^* = -\alpha^*W^*B\hat{S}J\beta - \dot{\beta} = -\alpha^*W_U^*B_{UL}(I - \hat{A}_{LL})^{-1}J_L\beta - \dot{\beta}$$

When there are no low-class people, $\hat{S}J = 0$ so $\dot{\alpha}^* + \dot{\beta} = 0$.

The main eigenvectors of a society with $q$ upper-class families are

$$V = \hat{V} + \varepsilon\hat{S}(B\hat{V} - \hat{V}\dot{\lambda}) + \varepsilon\hat{V}\dot{\beta} + O(\varepsilon^2)$$

$$U^* = \hat{U}^* + \varepsilon\hat{U}^*B\hat{S} + \varepsilon\dot{\alpha}^*\hat{U}^* + O(\varepsilon^2)$$

$\hat{V}\dot{\lambda} = J\beta\dot{\lambda} = J(W^*BJ)\beta = JW^*B\hat{V}$. Both equations can be re-written as

$$V = (I + \varepsilon \hat{S}(I - JW^*)B)\hat{V}(I + \varepsilon \dot{\beta}) + O(\varepsilon^2), \hat{V} = J\beta$$

(5)

$$U^* = (I + \varepsilon \dot{\alpha}^*)\hat{U}^*(I + \varepsilon B\hat{S}) + O(\varepsilon^2), \hat{U}^* = \alpha^* W^*$$

(6)

$I + \varepsilon \hat{S}(I - JW^*)B$ breaks the step-function structure of $V$ for upper-class families.

## III. Identify Families from Main Eigenvalues/Eigenvectors

The number of upper-class families in a society is the number of eigenvalues which are "very close" to 1 – including 1 itself. Unfortunately, an eigenvalue/eigenvector doesn't represent an upper-class family. For example, the trivial eigenvalue 1, trivial eigenvectors $v_1 = \eta/\sqrt{n}$ and $u_1^* = \sqrt{n} \cdot \omega^*$ – they represent all the families (in fact the whole society), not a specific one.

An algorithm to identify $q$ upper-class families from $q$ main eigenvalues/eigenvectors is proposed in this section. As an example, this algorithm is applied to a society with 50 members to demonstrate how it works.

Look at the politics matrix of this society (Figure 1):

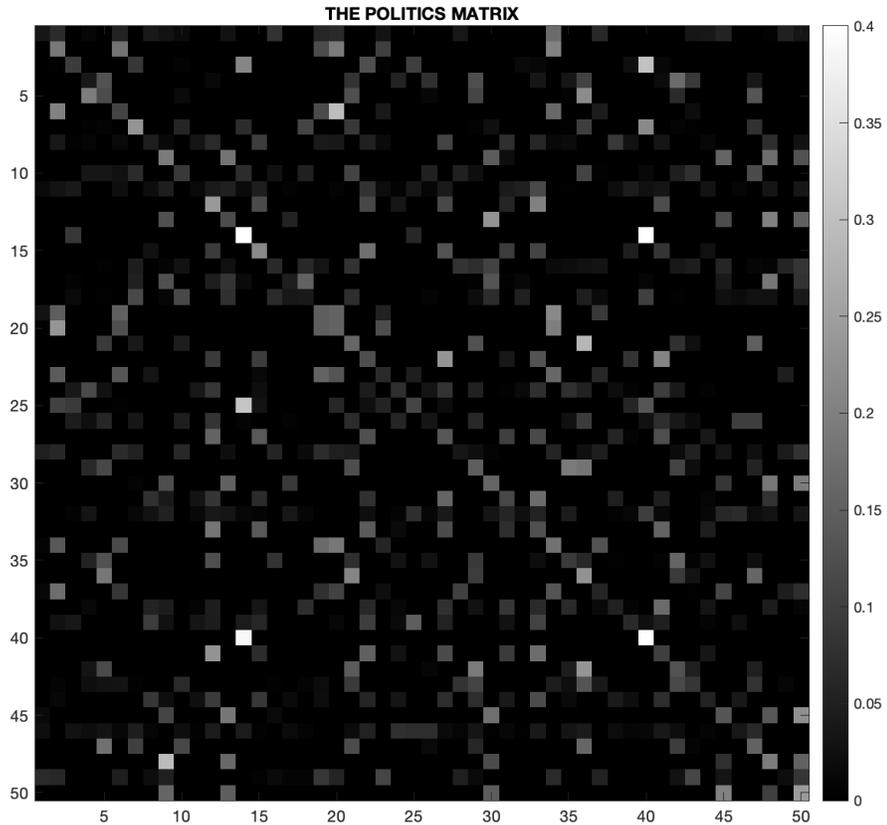

Figure 1

There are five eigenvalues near 1 – including 1 itself (Figure 2). So there should be five upper-class families which are not really "visible" in Figure 1.

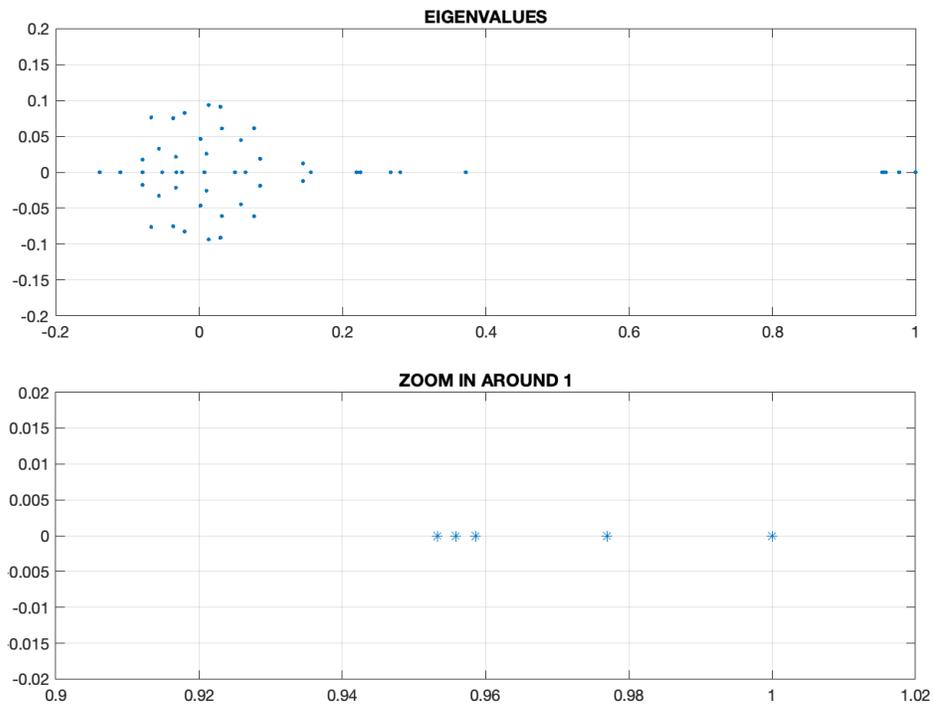

Figure 2

$V = J\beta + O(\varepsilon)$.

The first approach is to take the eigenvector $v_2$ corresponding to the 2$^{nd}$ largest eigenvalue $\lambda_2 = 0.977$, sort the components of this vector and expect people in the same upper-class family will group together because $v_2$ is almost like a step function for upper-class people. See Figure 3:

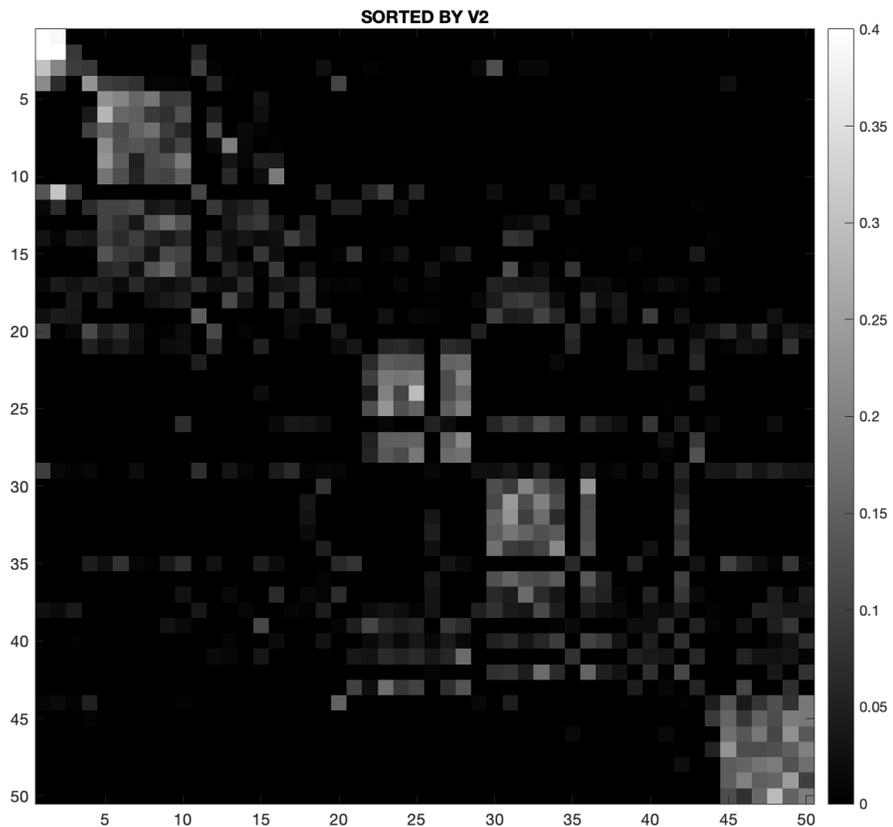

Figure 3

Those upper-class families show up roughly but not clearly. Probably it should work well if there are almost no low-class people. When $V_U = J_U \beta + O(\varepsilon)$ looks like a step function with $q$ steps, $V_L = \widehat{D}_{LU} J_U \beta + O(\varepsilon)$ doesn't. Low-class people appear anywhere when all the people are sorted by an eigenvector. Look at the five main eigenvectors in Figure 4 (sorted by $v_2$):

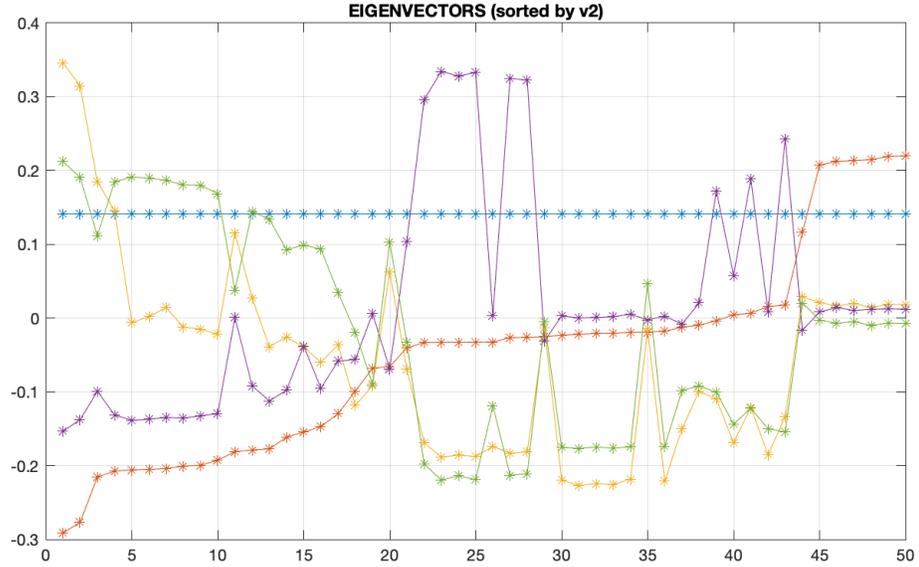

Figure 4

Five steps are supposed to be observed in the plot of sorted $v_2$ but only two steps are clear: the first two persons and the last six persons. Low-class people can be anywhere so it's hard to identify all upper-class families from one eigenvector $v_2$. For example, the 26th person in this plot doesn't really belong to "that family" – it can be seen from another three non-trivial eigenvectors. It's also clear in Figure 3, too.

When a column of $V$ doesn't represent an upper-class family, a column of $J$ does:

$$J = \begin{bmatrix} \eta_1 & \cdots & 0 \\ \vdots & \ddots & \vdots \\ 0 & \cdots & \eta_q \\ \widehat{D}_{L1}\eta_1 & \cdots & \widehat{D}_{Lq}\eta_q \end{bmatrix}$$

The $j$th column indicates how much people are "with" the $j$th upper-class family $U_j$. A component of this column is 1 for people in $U_j$, 0 for people in other upper-class families and it's between 0 and 1 for low-class people. Basically low-class people are people who are not with an upper-class family totally. $J$ is a partition of unity: $J \boxed{\eta} = \eta$.

Because $V = J\beta + O(\varepsilon)$, $J$ can be recovered roughly by $J = V\alpha^* + O(\varepsilon)$ if $\beta$ or $\alpha^*$ can be obtained from $V$, $U^*$ and $\lambda$. For any n-by-n diagonal matrix $\Lambda$, let's look at this matrix:

$$Z \equiv U^* \Lambda V = \alpha^* W^* \Lambda J \beta + O(\varepsilon) = \alpha^* \Omega \beta + O(\varepsilon)$$

$\Omega = W^* \Lambda J$ is an q-by-q diagonal matrix.

$$Z\alpha^* = \alpha^*\Omega + O(\varepsilon), \beta Z = \Omega\beta + O(\varepsilon)$$

So $\beta$ and $\alpha^*$ represent the eigenvectors of $Z$ – up to the error term $O(\varepsilon)$. Thus $J$ can be recovered roughly by the eigenvectors of $Z = U^*\Lambda V$.

Next step is to find a "good" $\Lambda$. $\Lambda$ cannot be the identity matrix. In this case $Z = U^*V = I$ and any vector is an eigenvector of $Z$.

The next-order terms of $V$ and $U^*$ in Equations (5) and (6) can help:

$$Z = (I + \varepsilon\dot{\alpha}^*)\alpha^*W^*(I + \varepsilon B\hat{S})\Lambda(I + \varepsilon\hat{S}(I - JW^*)B)J\beta(I + \varepsilon\dot{\beta}) + O(\varepsilon^2)$$

$$Z = \tilde{\alpha}^*\tilde{\Omega}\tilde{\beta} + O(\varepsilon^2)$$

$$\tilde{\Omega} = \Omega + \varepsilon W^*B\hat{S}\Lambda J + \varepsilon W^*\Lambda\hat{S}(I - JW^*)BJ$$

$$\tilde{\alpha}^* = (I + \varepsilon\dot{\alpha}^*)\alpha^*, \tilde{\beta} = \beta(I + \varepsilon\dot{\beta})$$

Let's try to find a suitable $\Lambda$ that $\tilde{\Omega}$ can be close to the diagonal matrix $\Omega = W^*\Lambda J$ as possible. $W^*\hat{S} = 0$ but $\Lambda$ cannot be $I$. However, $\varpi_j^*\hat{S}_j = 0$ for any $j$. $W^*\Lambda\hat{S} = 0$ when $\Lambda$ is in this form:

$$\Lambda = \begin{bmatrix} \Lambda_U & 0 \\ 0 & \Lambda_L \end{bmatrix}$$

$$\Lambda_U = \begin{bmatrix} c_1 I_1 & \cdots & 0 \\ \vdots & \ddots & \vdots \\ 0 & \cdots & c_q I_q \end{bmatrix}$$

(7)

$$W^*\Lambda\hat{S} = \begin{bmatrix} c_1\varpi_1^* & \cdots & 0 \\ \vdots & \ddots & \vdots \\ 0 & \cdots & c_q\varpi_q^* \end{bmatrix} \begin{bmatrix} \hat{S}_1 & \cdots & 0 \\ \vdots & \ddots & \vdots \\ 0 & \cdots & \hat{S}_q \end{bmatrix} = 0$$

Now $\tilde{\Omega} = \Omega + \varepsilon W^*B\hat{S}\Lambda J$. Because $\hat{S}_j\eta_j = 0$ for any $j$, $\hat{S}_U\Lambda_U J_U = 0$,

$$\hat{S}\Lambda J = \begin{bmatrix} \hat{S}_U & 0 \\ \hat{D}_{LU}\hat{S}_U & (I - \hat{A}_{LL})^{-1} \end{bmatrix} \begin{bmatrix} \Lambda_U J_U \\ \Lambda_L J_L \end{bmatrix} = \begin{bmatrix} 0 \\ (I - \hat{A}_{LL})^{-1}\Lambda_L J_L \end{bmatrix}$$

$$W^*B\hat{S}\Lambda J = W_U^*B_{UL}(I - \hat{A}_{LL})^{-1}\Lambda_L J_L$$

Let $\breve{\beta}$ ($\approx \beta$ or $\tilde{\beta}$) be the q-by-q diagonal matrix representing all left eigenvectors of $Z$:

$$\breve{\beta}Z = \left(\Omega + \varepsilon\dot{\Omega} + O(\varepsilon^2)\right)\breve{\beta}$$

$\breve{J} \equiv V\breve{\beta}^{-1}$ would be the approximation of $J$ to identify $q$ upper-class families of this society:

$$\breve{J} = (I + \varepsilon\hat{S}(I - JW^*)B)J\tilde{\beta}\breve{\beta}^{-1} + O(\varepsilon^2)$$

$I + \varepsilon\hat{S}(I - JW^*)B$ is "out of control" but it would be great if $\breve{\beta} = \tilde{\beta} + O(\varepsilon^2)$. Let's assume

$$\breve{\beta} = (I + \varepsilon x)\tilde{\beta} + O(\varepsilon^2)$$

$$(I + \varepsilon x)\tilde{\beta}\tilde{\alpha}^*\left(\Omega + \varepsilon W_U^* B_{UL}(I - \hat{A}_{LL})^{-1}\Lambda_L J_L\right)\tilde{\beta} = (\Omega + \varepsilon\dot{\Omega})(I + \varepsilon x)\tilde{\beta} + O(\varepsilon^2)$$

$$\tilde{\beta}\tilde{\alpha}^* = \beta(I + \varepsilon\dot{\beta} + \varepsilon\dot{\alpha}^*)\alpha^* + O(\varepsilon^2) = I - \varepsilon W_U^* B_{UL}(I - \hat{A}_{LL})^{-1} J_L + O(\varepsilon^2)$$

The equation for $x$ is

$$x\Omega - W_U^* B_{UL}(I - \hat{A}_{LL})^{-1} J_L\Omega + W_U^* B_{UL}(I - \hat{A}_{LL})^{-1}\Lambda_L J_L = \dot{\Omega} + \Omega x$$

$$W_U^* B_{UL}(I - \hat{A}_{LL})^{-1}(\Lambda_L J_L - J_L\Omega) = \dot{\Omega} + \Omega x - x\Omega$$

$\Lambda_L$ is an l-by-l diagonal matrix and $\Omega$ is an q-by-q diagonal matrix. Unfortunately $\Lambda_L J_L - J_L\Omega$ cannot be zero except $\Lambda_L = cI_{l\times l}$ and $\Omega = cI_{q\times q}$. But $\Omega = W^*\Lambda J$ cannot be like this.

So, the best shot is to find a diagonal $\Lambda$ in a form of Equation (7) as possible – don't forget those $q$ upper-class families are unknown. $V = J\beta + O(\varepsilon)$, every eigenvector is almost like a step function for upper-class people. So $\Lambda = diag(v_j)$ is a good choice for any $j$ except $v_1 = \eta/\sqrt{n}$.

Another candidate can be found in this way – and it's probably better:

$$VU^* = JW^* + O(\varepsilon) = \begin{bmatrix} J_U W_U^* & 0 \\ J_L W_U^* & 0 \end{bmatrix} + O(\varepsilon)$$

$$J_U W_U^* = \begin{bmatrix} \eta_1\varpi_1^* & \cdots & 0 \\ \vdots & \ddots & \vdots \\ 0 & \cdots & \eta_q\varpi_q^* \end{bmatrix} + O(\varepsilon)$$

The diagonal of $VU^*$ is not like a step function for upper-class people. However,

$$u_1^* = \sqrt{n}\cdot\omega^* = \alpha_1^* W^* + O(\varepsilon) = [\alpha_{11}^*\varpi_1^*\ \alpha_{12}^*\varpi_2^*\ \ldots\ \alpha_{1q}^*\varpi_q^*\ 0_{1\times l}] + O(\varepsilon)$$

$$\Lambda \equiv (diag(u_1^*))^{-1} diag(VU^*)$$

$$\Lambda_U = \begin{bmatrix} \frac{I_1}{\alpha_{11}^*} & \cdots & 0 \\ \vdots & \ddots & \vdots \\ 0 & \cdots & \frac{I_q}{\alpha_{1q}^*} \end{bmatrix} + O(\varepsilon)$$

Because $u_1^* = \sqrt{n} \cdot \omega^* > 0$, this $\Lambda$ is well-defined.

$\breve{J} \equiv V\breve{\beta}^{-1} = J + O(\varepsilon)$ is the n-by-q matrix to identify $q$ "families" – the maximum of a row is the family that person belongs to. In this case a low-class person is supposed to be included in the family he's with most and an upper-class person will be in the family he belongs to as well. In fact the separation between upper-class and low-class people is not strict – expect $\varepsilon$ is as small as an infinitesimal in the era of Newton and Leibniz.

**The Algorithm**

1. Find $q$ eigenvectors close to 1 and the corresponding eigenvectors that
$$AV = V\lambda, U^*A = \lambda U^*, I = diag(V^*V) = U^*V$$
2. $\Lambda \equiv \left(diag(u_1^*)\right)^{-1} diag(VU^*)$ and $Z \equiv U^*\Lambda V$.
3. Find all the left eigenvectors of $Z$ as an q-by-q matrix $\breve{\beta}$. $\breve{\beta}$ is normalized that the first column is $\boxed{\eta}/\sqrt{n}$ – just like $\beta$.
4. The n-by-q matrix $\breve{J} \equiv V\breve{\beta}^{-1}$ is an approximation of $J$ which indicates how much each individual is with each upper-class family roughly.
5. $\breve{J}$ divides the society $\{1,2,\ldots n\}$ into $q$ disjoint subsets $\breve{U}_1, \breve{U}_2, \ldots \breve{U}_q$. For any $j$, an upper-class family $U_j$ is supposed to be inside $\breve{U}_j$.

Let's sort all the people based on $\breve{J}$ to make those upper-class families as "visible" as possible. Look at this q-by-q matrix:

$$J^*J = J_U^*J_U + J_L^*J_L = \begin{bmatrix} m_1 & \cdots & 0 \\ \vdots & \ddots & \vdots \\ 0 & \cdots & m_q \end{bmatrix} + J_L^*J_L$$

The l-by-q matrix $J_L$ indicates how much every low-class person is with each upper-class family. So $(J^*J)_{ij} = (J_L^*J_L)_{ij}$ indicates how much low-class people are with both $U_i$ and $U_j$ "simultaneously" when $i \neq j$ – or the strength $U_i$ and $U_j$ are "linked" through low-class people.

$\breve{J}^*\breve{J}$ is an approximation of $J^*J$. Let's apply the algorithm to that society with 50 members:

$$\check{J}^*\check{J} = \begin{bmatrix} *.** & 0.57 & 0.64 & 0.66 & 0.25 \\ 0.57 & *.** & 0.48 & 0.32 & 0.21 \\ 0.64 & 0.48 & *.** & 0.94 & 0.92 \\ 0.66 & 0.32 & 0.94 & *.** & 0.30 \\ 0.25 & 0.21 & 0.92 & 0.30 & *.** \end{bmatrix}$$

When $\breve{U}_j$ includes low-class people who are with $U_j$ most, let's try to put "families" next to each other as possible when they are linked through low-class people more. $\breve{U}_3$ and $\breve{U}_4$ are linked most (0.94). Then try to find $\breve{U}_j$ which are linked to $\breve{U}_3$ or $\breve{U}_4$ most – it's $\breve{U}_5$ (0.92 to $\breve{U}_3$) and this three subsets are sorted in $\breve{U}_5 - \breve{U}_3 - \breve{U}_4$. Find next $\breve{U}_j$ which are linked to $\breve{U}_5$ or $\breve{U}_4$ most… eventually these five subsets are sorted in this way:

$$\breve{U}_5 - \breve{U}_3 - \breve{U}_4 - \breve{U}_1 - \breve{U}_2$$

Now let's sort the members inside each $\breve{U}_j$. For a subset with two "neighbors" – for example $\breve{U}_3$, it can be separated into two disjoint subsets – people with $U_5$ more ($\breve{U}_3^5$) and people with $U_4$ more ($\breve{U}_3^4$) (based on $\check{J}$)… Now there are eight subsets sorted in this way:

$$\breve{U}_5 - (\breve{U}_3^5 - \breve{U}_3^4) - (\breve{U}_4^3 - \breve{U}_4^1) - (\breve{U}_1^4 - \breve{U}_1^2) - \breve{U}_2$$

For subset $\breve{U}_3$, people with $U_5$ more can be close to $U_5$ and people with $U_4$ more can be close to $U_4$ as well.

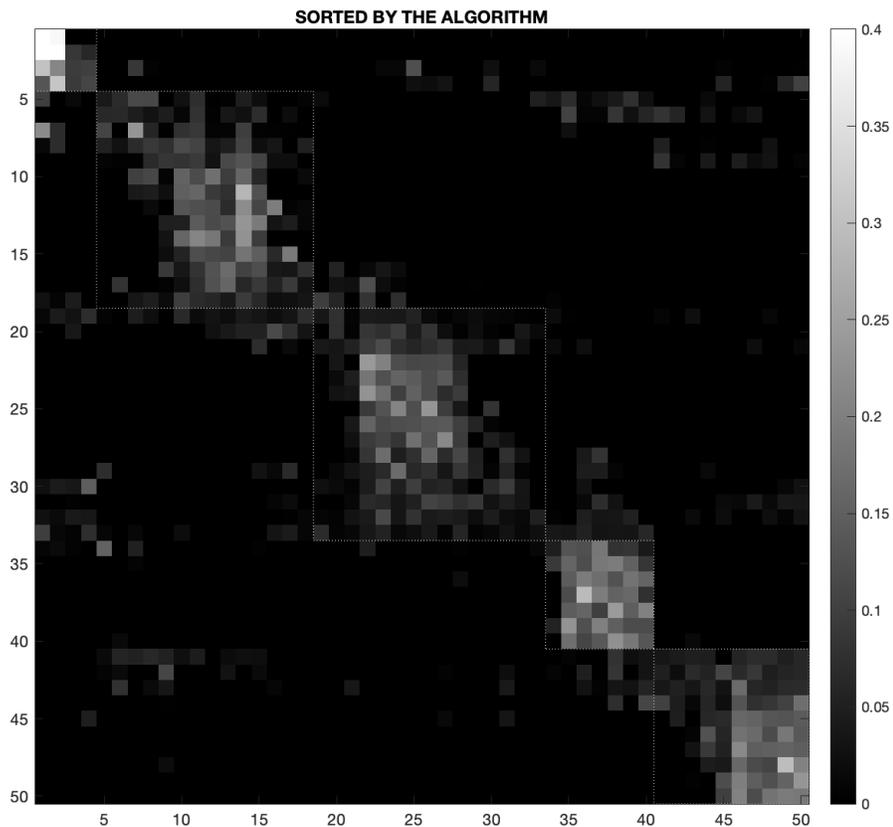

Figure 5

The final step is to sort members inside these eight subsets. Sort both $\breve{U}_3^5$ and $\breve{U}_3^4$ by the 3rd column of $\check{J}$ separately that people with $U_3$ more are on the right side of $\breve{U}_3^5$ and the left side of $\breve{U}_3^4$.

In this case $U_3$ is supposed to be in

the middle of $(\breve{U}_3^5 - \breve{U}_3^4)$, "protected" by low-class people who are with $U_3$ most and $U_5$ more on the left side, low-class people who are with $U_3$ most and $U_4$ more on the right side.

$(\breve{U}_4^3 - \breve{U}_4^1)$ and $(\breve{U}_1^4 - \breve{U}_1^2)$ are sorted in the same way. $\breve{U}_5$ is sorted by the 5th column of $\breve{J}$ that people with $U_5$ more are on the left side and $\breve{U}_2$ is sorted by the 2nd column of $\breve{J}$ that people with $U_2$ more are on the right side. See Figure 5: five dotted squares are the disjoint subsets defined by $\breve{J}$ and five upper-class families are inside these dotted squares.

Figure 6 shows five columns of $\breve{J}$ – see how they define five families of this society.

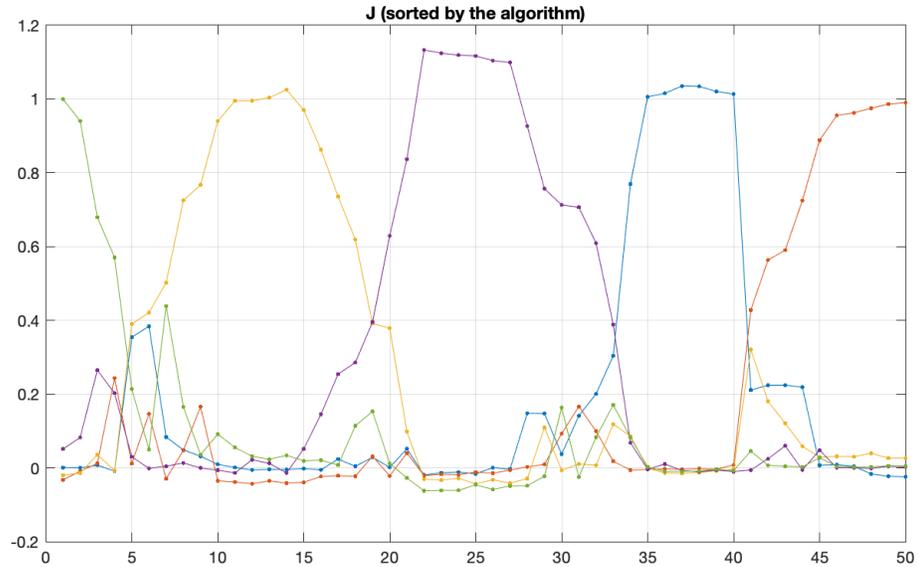

**Figure 6**

Figure 7 shows the power $\omega^*$ of this society. Upper-class people get meaningful power – and there are five upper-class families.

The algorithm works fine in this sample society.

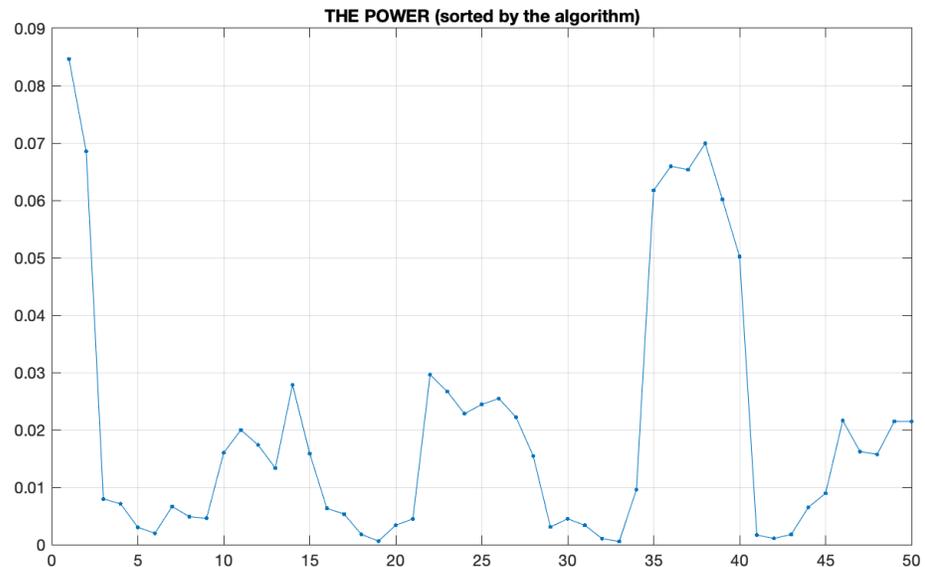

**Figure 7**

## IV. Application

The motivation for people to form families has been shown in [1]. To be with a family might not be enough to survive in a society – it's also important to identify other families. Furthermore, someone can even be in a family without knowing the leader and most people in this family. It can be important for him to identify his own family as well.

Of course most times families are identified based on instinct. The most famous "family" identified in 21$^{st}$ century is probably the "Axis of Evil". This family was identified by President George W. Bush in 2002 – comprising Iran, Iraq and North Korea. However, it's not based on any reality. Iran and Iraq fought for eight years in 1980s and it's hard to believe they were linked in 2002. North Korea was not even with China totally. Why was this country with Iran and Iraq?

Another famous example is the "Gang of Four" who came to prominence during the Cultural Revolution (1966-1976). Chairman Mao was the first one to use this term in 1974. He made some strong criticisms of the four – his third wife included. However, Mao took no action against them. As a political genius, Mao identified this gang accurately. They were arrested, tried and imprisoned after Mao's death in 1976.

The algorithm developed in this paper offers a possible scientific way to identify families in a society.